# Design and Realization of an *S*-Band Microwave Low-Noise Amplifier for Wireless RF Subsystems

Ardavan Rahimian[1], Davood Momeni Pakdehi[2]

*Abstract*—**This study undertakes the theoretical design, CAD modeling, realization, and performance analysis of a microwave low-noise amplifier (LNA) which has been accurately developed for operation at 3.0 GHz (*S*-band). The objective of this research is to thoroughly analyze and develop a reliable microstrip LNA intended for a potential employment in wireless communication systems, and satellite applications. The *S*-band microwave LNA demonstrates the appropriateness to develop a high-performance and well-established device realization for wireless RF systems. The microwave amplifier simulations have been conducted using the latest version of the AWR Design Environment software.**

## I. INTRODUCTION

THE modern emerging of the high-speed and high-data rate wireless communications has encouraged intensive high-profile research in both the academic and industrial fields in order to develop high-performance RF/microwave systems for the potential employment in digital wireless communications. Radiowave propagation is very lossy, so the signals travelling from the transmitter normally suffer from degradation. When these signals are received at the receiver antenna, they are very weak. Low-noise amplifier (LNA), which is the combination of the low-noise, high-gain, and stability over the entire range of operating frequency and is placed very close to the antenna, is crucial RF component employed in communication systems in order to amplify these very weak RF signals captured by the antenna, and also to boost desired signal power while adding as little noise and distortion as possible so that the noise added by later components of the RF receiver chain has less effect on the signal-to-noise ratio (SNR), and the overall performance.

## II. MICROWAVE LNA: DESIGN AND ANALYSIS

The RF low-noise amplifier component design remains as one of the challenging tasks in the RF receiver system design, which plays an important role in the system noise performance or sensitivity of the total receiver chain; since it is mandatory to meet several crucial requirements, such as: low noise-figure (*NF*) to enhance the RF sensitivity, optimum gain to reduce the RF mixer noise, broadband input matching to improve the reflection coefficient, and reasonable efficiency for the low-power system consumption [1]. The signal is typically filtered, amplified by LNA, and translated to the baseband by mixing with a local oscillator. After being demodulated, the RF signal is then applied to an ADC system which digitizes the analog signal; and then this digital signal is processed in a DSP unit. The low-noise amplifier enhances the level of the RF signal incident on its input without introducing significant noise and distortion. As the first real signal processing element after the antenna system, and the sensitive block in the RF receiver, the LNA determines the noise and linearity performance of the overall microwave system as its primary role in the system [2]. The primary parameters for the high-performance LNA design are gain and system bandwidth, noise, linearity, impedance matching and power consumption. The *NF* must be minimized to provide high-gain with sufficient linearity and establishing the compatibility to other transceiver blocks (i.e. impedance matching) [3]. In the wireless system with a series of cascaded components, where each stage adds additional noise, the first stage and its noise and gain characteristics are critical. Friis's formula, as in (1), calculates the total noise factor of a system with cascaded stages, wherein the $F_1$ and $G_1$ are the RF noise factor and gain of the first stage, respectively; hence, the LNA supplies sufficient gain in order to overcome the noise of the succeeding stages while at the same time producing as little noise as possible itself. Also, as the amplifiers have gain, noise added in the later stages does not have as much of an impact as noise added in the first stage of the microwave receiver [4].

$$\xrightarrow{yields} F_{total} = F_1 + \frac{F_2 - 1}{G_1} + \frac{F_3 - 1}{G_1 G_2} + \frac{F_4 - 1}{G_1 G_2 G_3} + \cdots \quad (1)$$

In order to maximize the RF/microwave low-noise amplifier power transfer from source to load, matching impedances is required. As seen in fig. 1, in an RF circuit where the source and load impedances are fixed, the LNA matching networks are designed so that $Z_s$ matches $Z_I$, and $Z_L$ matches $Z_2$, at the RF matching networks input and output, respectively. In this case the maximum power transfer will occur when the reactive components of the impedances cancel each other, and that is when they are complex conjugates (conjugate matching). As the microwave amplifier operates with RF input signals of a particular frequency band it is desired to design the component with an RF center frequency and bandwidth, accordingly [5].

[1]School of Electronic, Electrical and Computer Engineering, University of Birmingham; Edgbaston, Birmingham B15 2TT, UK (rahimian@ieee.org)
[2]Faculty of Electrical Engineering, Bremen University of Applied Sciences; Bremen 28199, Germany (dmomeni-pakdehi@stud.hs-bremen.de)

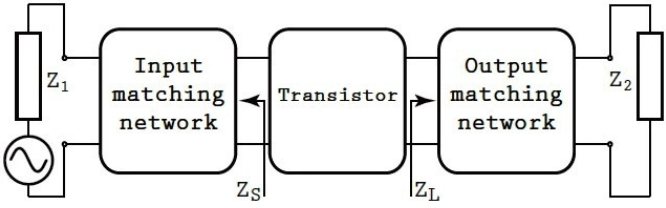

Fig. 1. The microwave LNA block diagram with matching networks [5].

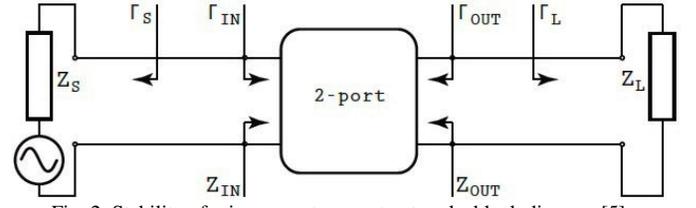

Fig. 2. Stability of microwave two-port networks block diagram [5].

In order to order to accurately design a well-established and high-performance RF low-noise component with an accurate matching network to transfer maximum power and to improve the SNR, selection of the appropriate transistor is of crucial importance for the microwave system design and realization. Specified *S*-parameters for the LNA modeling and simulations for the intended RF operating frequency should be taken into account in order to primarily check the amplifier stability (i.e. its resistance to oscillations), and also to potentially design the LNA matching networks to have an optimum compromise between the gain and the *NF*. Matching is employed in order to maximize SNR in the receiver input stages, and minimize signal distortion in the transmission lines. One of the primary steps in the LNA design process is the confirmation that the transistor is unilateral, and that satisfies equation (1), as given:

$$\xrightarrow{yields} \frac{1}{(1+U)^2} < \frac{G_T}{G_{TU}} < \frac{1}{(1-U)^2}. \qquad (2)$$

where $G_T$ and $G_{TU}$ are the transducer and unilateral transducer gain ($S_{12} = 0$), respectively; and $U$ is given as (3), as well [6]:

$$\rightarrow U = \frac{|S_{11}|.|S_{12}|.|S_{21}|.|S_{22}|}{(1-|S_{11}|^2)(1-|S_{22}|^2)}. \qquad (3)$$

The stability of the intended amplifier is a very important consideration in a design, and can be determined from the *S*-parameters, the matching networks, and the terminations. In a stability perspective, an LNA can be either unconditionally stable or potentially unstable; there are two possibilities, either stability factor, *k* is smaller or greater than 1. If *k* is greater than 1, the device is then unconditionally stable. There is no combination of passive source or load impedance that will cause the device to oscillate. If *k* is smaller than 1, the device is conditionally stable and therefore potentially unstable. It can be induced into oscillation by certain passive source and load impedances. In a 2-port network (fig. 2), oscillation may occur when some load and source termination cause the input and output impedance to have a negative real part. There are three main causes for the mentioned scenario: the internal feedback, external feedback, and also the excessive gain at out-of-band frequencies. In order to prevent instability, the aim is to place $\Gamma_S$ and $\Gamma_L$ in the stable region of the Smith chart. In practice, this is done with filtering and resistive loading to attenuate the gain [7]. The condition for the unconditional stability is as (4):

$$\xrightarrow{yields} K = \frac{1-|S_{11}|^2-|S_{22}|^2+|\Delta|^2}{2|S_{12}S_{21}|} > 1, \qquad (4)$$
$$where \; \Delta = S_{11}S_{22} - S_{12}S_{21}.$$

It is common practice to graphically illustrate the region for which the LNA is unconditionally stable with stability circles in the Smith chart. Fig. 3 explains the concept of the stability circles that includes the input stability circle in the $\Gamma_S$-plane, and output stability circle in the $\Gamma_L$-plane. The circumstances determine whether the stable region is inside or outside circle. Also, when the RF design specifications have been determined along with the active device and its working point, the LNA then initiates with the plots of source stability-, constant noise-, and constant gain-circles, based on the following obtained equations, to determine the $\Gamma_S$ and $\Gamma_L$. With the equation for noise in a 2-port microwave amplifier slightly altered with the system normalized noise resistance and source admittance, the equation (5) is further obtained, in which $y_s$ and $y_{opt}$ have been expressed in terms of reflection coefficients, as in (6), and (7).

$$\xrightarrow{yields} F = F_{min} + \frac{r_n}{g_s}|y_s - y_{opt}|^2. \qquad (5)$$
$$\rightarrow y_s = \frac{1-\Gamma_s}{1+\Gamma_s}; \qquad (6)$$
$$\rightarrow y_{opt} = \frac{1-\Gamma_{opt}}{1+\Gamma_{opt}}. \qquad (7)$$

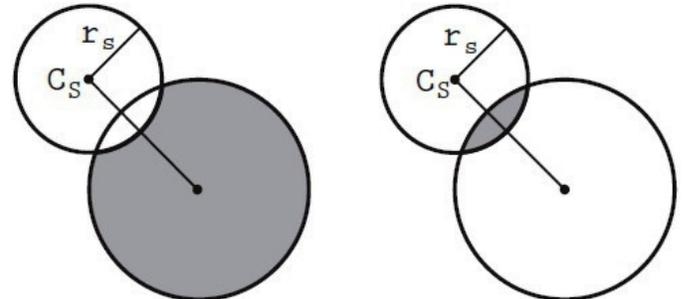

Fig. 3. a) Smith chart presenting grey stable region in the $\Gamma_L$ plane; left: $|S_{11}| < 1$; right: $|S_{11}| > 1$; taken from [5].

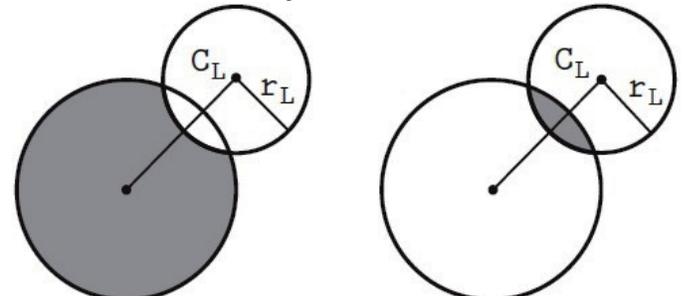

Fig. 3. b) Smith chart presenting grey stable region in the $\Gamma_S$ plane; left: $|S_{22}| < 1$; right: $|S_{22}| > 1$; taken from [5].

These relations yield the following expression, as in (8):

$$\xrightarrow{yields} F = F_{min} + \frac{4r_n|\Gamma_s - \Gamma_{opt}|^2}{(1 - |\Gamma_s|^2)|1 + \Gamma_{opt}|^2}. \tag{8}$$

Equation (8) shows that the *NF* depends on the variable $\Gamma_s$ and three quantities know as the noise parameters, which are given by the transistor manufacturer, including: $F_{min}$, $r_n$, and $\Gamma_{opt}$. For a specified *NF* $F_i$ the equation becomes then as (9), with the center and radius of the circle as in (10) and (11), respectively.

$$\xrightarrow{yields} \left|\Gamma_s - \frac{\Gamma_{opt}}{1 + N_i}\right|^2 = \frac{N_i^2 + N_i(1 - |\Gamma_{opt}|^2)}{(1 + N_i)^2}. \tag{9}$$

$$\xrightarrow{center} C_{Fi} = \frac{\Gamma_{opt}}{1 + N_i}; \tag{10}$$

$$\xrightarrow{radius} r_{Fi} = \frac{1}{1 + N_i}\sqrt{N_i^2 + N_i(1 - |\Gamma_{opt}|^2)}. \tag{11}$$

The mathematically-oriented derivations for the constant available gain circles have been determined, as expressed with equation (12), in which the center and radius of the circle are also given as equation (13), and equation (14), respectively.

$$\xrightarrow{yields} G_A = |S_{21}|^2 \frac{1 - |\Gamma_s|^2}{\left(1 - \left|\frac{S_{22} - \Delta\Gamma_s}{1 - S_{11}\Gamma_s}\right|^2\right)|1 - S_{11}\Gamma_s|^2} \tag{12}$$

$$= |S_{21}|^2 g_a.$$

$$\xrightarrow{center} C_a = \frac{g_a C_1^*}{1 + g_a(|S_{11}|^2 - |\Delta|^2)}, \tag{13}$$

$$where\ C_1 = S_{11} - \Delta S_{22}^*;$$

$$\xrightarrow{radius} r_a = \frac{\sqrt{1 - 2K|S_{12}S_{21}|g_a + |S_{12}S_{21}|^2 g_a^2}}{|1 + g_a(|S_{11}|^2 - |\Delta|^2)|}. \tag{14}$$

Conclusively, the system output- and input-stability have been plotted based on the simulations with the relations as in equations (15) and (16), along with the specified centers and radii as equation (17) to equation (20), respectively [5, 8, 9].

$$\xrightarrow{yields} \left|\Gamma_L - \frac{(S_{22} - \Delta S_{11}^*)^*}{|S_{22}|^2 - |\Delta|^2}\right| = \left|\frac{S_{12}S_{21}}{|S_{22}|^2 - |\Delta|^2}\right|, |\Gamma_{IN}| = 1; \tag{15}$$

$$\xrightarrow{yields} \left|\Gamma_L - \frac{(S_{11} - \Delta S_{22}^*)^*}{|S_{11}|^2 - |\Delta|^2}\right| = \left|\frac{S_{12}S_{21}}{|S_{11}|^2 - |\Delta|^2}\right|, |\Gamma_{OUT}| = 1. \tag{16}$$

$$\xrightarrow{center\ (L)} C_L = \frac{(S_{22} - \Delta S_{11}^*)^*}{|S_{22}|^2 - |\Delta|^2}; \tag{17}$$

$$\xrightarrow{center\ (S)} C_S = \frac{(S_{11} - \Delta S_{22}^*)^*}{|S_{11}|^2 - |\Delta|^2}; \tag{18}$$

$$\xrightarrow{radius\ (L)} r_L = \left|\frac{S_{12}S_{21}}{|S_{22}|^2 - |\Delta|^2}\right|; \tag{19}$$

$$\xrightarrow{radius\ (S)} r_S = \left|\frac{S_{12}S_{21}}{|S_{11}|^2 - |\Delta|^2}\right|. \tag{20}$$

When $\Gamma_s$ is selected, the proceedings continue by determining the output reflection coefficient ($\Gamma_{out}$) of the microwave LNA, and then plot load stability circles. If then $\Gamma^*_{out}$ is in the stable region, $\Gamma_L$ is set to $\Gamma^*_{out}$ for a complex conjugate matched RF output. Should that not be the case, transducer gain circles are drawn to find a $\Gamma_L$ in the stable region that leads to reasonably high-transducer gain. The procedures based on the specified steps for the accurate design of the RF low-noise amplifier for operation at 3.00 GHz have been conducted. A MATLAB-based script has been written in order to accurately perform the calculations, and to obtain the required design parameters for the various RF system transmission lines. Based on these obtained values, and the designed source and load reflection coefficients, matching networks at the source- and load-side have been designed, such that the gain is high, *NF* is 0 dB, and the stability is also maintained for a wide RF operating band.

## III. MICROWAVE LNA: SIMULATION AND REALIZATION

The design and simulation procedures have been thoroughly conducted in order to accurately realize the *S*-band microwave LNA. The layout is the view of the physical representation of schematic which is of crucial importance of the design (fig. 4), since the response of the circuit is dependent on the geometric shapes with which it is composed. To fabricate the LNA and to measure the realized system performance (fig. 5), the first step is to generate the layout from the schematic, by removing all the wires and lumped elements in the ideal system design, and connecting the RF components using the microstrip lines.

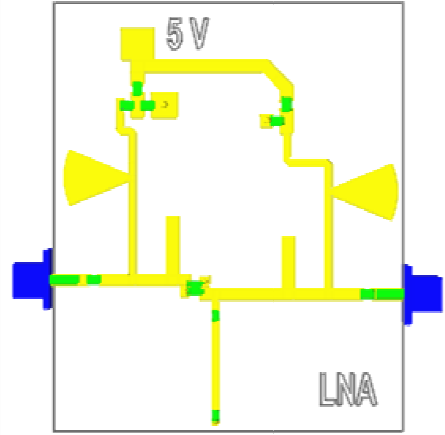

Fig. 4. *S*-band microwave LNA simulated layout.

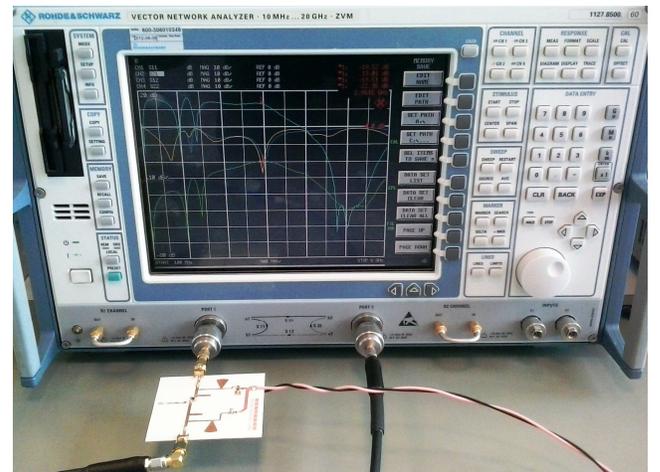

Fig. 5. *S*-band microwave LNA RF measurement setup.

As fig. 4 shows, the biasing circuit has also been developed separately, to set appropriate biasing conditions. The biasing circuit provides the desired voltage level for the intended RF performance, based on a low DC resistance in order to ensure that the microwave circuit is not loaded, and signals do not flow onto the supply lines; based on using a radial stub after the $\lambda$/4 high impedance biasing line, which helps to achieve the proper isolation at the desired frequency. For the transistor to be unconditionally stable as well, both the input and output stability circles must be completely outside the Smith chart; hence, using the software, stability circles have been plotted to check stability at the operating frequency using the frequency sweep of intended frequency range. The resulted plot satisfies the condition of the unconditional stability based on the $\mu$-test (fig. 6). The employed transistor is the Infineon Technologies AG N420, with the *S*-parameters as: $S_{11}$: 0.499<151.5°; $S_{21}$: 4.426<51.4°; $S_{12}$: 0.084<37.3°; and $S_{22}$: 0.161<-120.6°; hence, the parameters including: $\Delta$ = 0.336<-79.6°; $\Gamma_S$ = 0.697<-157°; and $\Gamma_L$ = 0.516<85° have been further obtained, in order to thoroughly design the amplifier components. Also, as fig. 7 presents the simulated frequency response confirms the device high-performance operation. Fig. 9 also presents the fabricated LNA (fig. 8) measured response. A comparison has been made with the simulated response; it has been shown that there is a good agreement between simulated and experimental results.

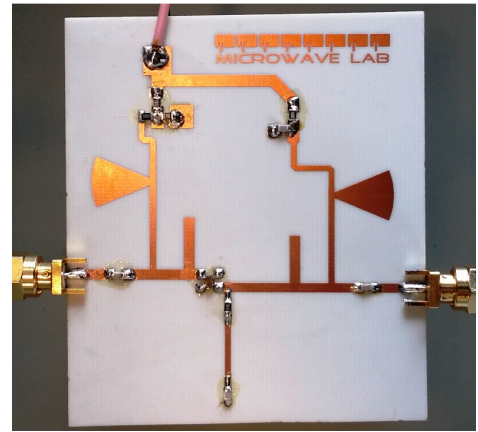

Fig. 8. Fabricated *S*-band microwave LNA.

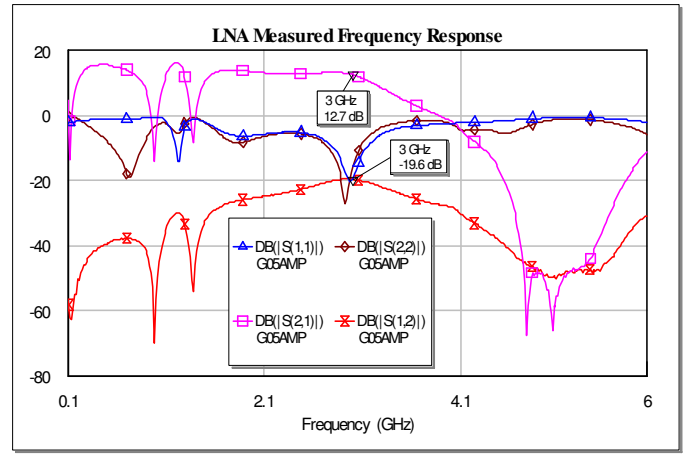

Fig. 9. *S*-band microwave LNA measured frequency response.

## IV. CONCLUSION

In this contribution, the procedures in order to accurately design, analysis, fabricate, and measure the novel and high-performance *S*-band microwave LNA have been carried out, for the potential employment in the wireless RF subsystems.

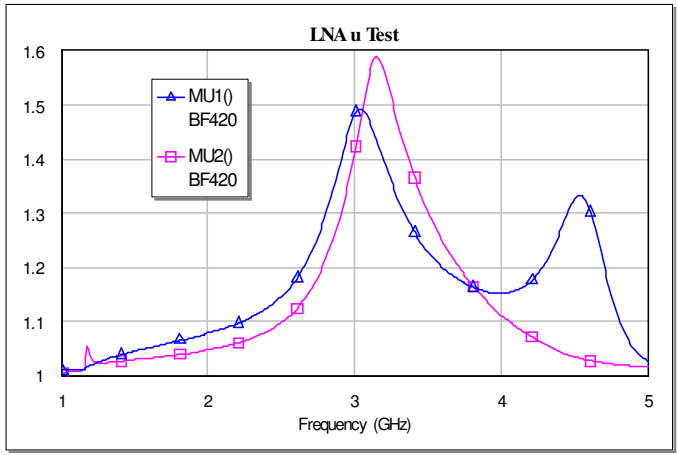

Fig. 6. *S*-band microwave LNA simulated stability $\mu$-test.

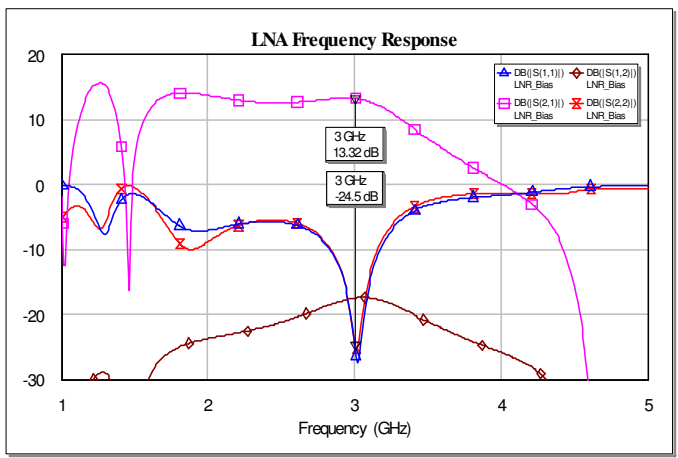

Fig. 7. *S*-band microwave LNA simulated frequency response.